\documentclass{article}

\usepackage{arxiv}
\usepackage{booktabs}
\usepackage{siunitx}
\usepackage[utf8]{inputenc} 
\usepackage[T1]{fontenc}    
\usepackage{hyperref}       
\usepackage{url}            
\usepackage{booktabs}       
\usepackage{amsfonts}       
\usepackage{nicefrac}       
\usepackage{microtype}      
\usepackage{graphicx}
\usepackage[square,compress,numbers]{natbib}
\usepackage{doi}
\usepackage{amssymb}
\usepackage{amsmath}
\usepackage{float}
\usepackage{lineno}

\title{From Logistic to Gompertz: A Microscopic Theory of Coherent Growth in Biological Systems}

\author{Matz A. Haugen, \\
Independent Researcher, Palo Alto, CA, USA\\
\texttt{matzhaugen@gmail.com} \\
  \And
  Dorothea Gilbert \\
  Independent Researcher, Oslo, Norway\\
}

\renewcommand{\shorttitle}{On the nature of the Gompertz}

\hypersetup{
pdftitle={From Logistic to Gompertz: A Microscopic Theory of Coherent Growth in Biological Systems},
pdfsubject={q-bio.NC, q-bio.QM},
pdfauthor={},
pdfkeywords={},
}

\begin{document}
\maketitle

\begin{abstract}
Logistic and Gompertz growth have traditionally been connected by augmenting the logistic model with an extra parameter, giving $\theta$-logistic or Richards growth. In spite of this bridge, the biological foundation of Gompertz growth remains only vaguely understood.
We propose a novel microscopic version of the Richards model where a coherence parameter sets the coupling between nodes on a network.
This model reveals Gompertz growth ($\theta \to 0$) as the coherent limit within this family: the system is asymptotically stable with a spectral gap protecting the macroscopic state, and each entity contributes linearly to the collective growth rate regardless of network topology.
In contrast, Richards ($\theta > 0$) and logistic ($\theta = 1$) growth require synchronization as a precondition for macroscopic validity, a condition that Gompertz growth instead imposes.
The coherence parameter $\theta$ thus acts as a symmetry-breaking parameter: at $\theta = 0$ the aggregate dynamics depend only on the collective mean and are invariant to how fluctuations are distributed among entities, an invariance broken at first order in $\theta$, where the macroscopic drift acquires a dependence on the microscopic variance.
These observations support interpreting Gompertz growth in biological systems as driven by a source external to the individual entities: a field that stimulates a response simultaneously across all entities, such as an environmental stressor, an electromagnetic field, or time over longer horizons.
Our results also admit a phenomenological classification of well-known growth models: uncorrelated growth without time dependence yields the exponential function, uncorrelated growth with linear time dependence yields the Gaussian, pairwise correlated growth yields the logistic, and correlated growth (or independent growth with a common driver) yields the Gompertz.
\end{abstract}


\section*{Highlights}
\begin{itemize}
\item Logistic and Gompertz growth is connected within a novel framework based on microscopic principles.
\item Gompertz growth is connected with a ubiquitous growth-activating field.
\item The Gompertz model exhibits properties not exhibited by the logistic function, namely closure under aggregation, mean-variance decoupling, macroscopic stability, and linear forcing from each microscopic entity.
\item Growth processes are put in a larger contextual framework based on varying dependence in spacetime, including the Exponential, Gaussian distributions. 
\end{itemize}

\section{Introduction}

In 1825, Gompertz published a paper on the nature of mortality where he discovered that from adulthood and onward ``the power to avoid destruction'' decreases exponentially with age. Equivalently, mortality rates increase in geometric progression for arithmetic increases in age, an observation now commonly referred to as Gompertz' law of mortality \cite{Gompertz1825}. Makeham (1889) later provided an interpretation to Gompertz' law where the recuperative force of a human, or ``vital force'', becomes less efficient with time as a ``stressor'' \cite{makeham1889further}. Gompertz' law has later been known as the Gompertz model or Gompertz growth. 

\subsection{Gompertz' law in existing work}
This simple law has been shown to generalize to situations where mortality rates are substituted by many types of abundance variables, like the growth of animals, plants, bacteria, and cancer tumors \cite{weymouth1931relative,weymouth1931age,laird1964dynamics,zwietering1990modeling,skinner1994mathematical,starck1998avian,aggrey2002comparison,paine2012fit,benzekry2014classical,halmi2014evaluation,tjorve2010shapes}, exposing Gompertz growth as nearly ubiquitous in biological growth processes. The idea that the Gompertz' law can be applied to other biological systems was first proposed by Weymouth (1931) \cite{weymouth1931relative,weymouth1931age} and Winsor (1932) \cite{Winsor1932}. Gompertz' law has also been shown to apply to societal phenomena like growth in railway traffic, and the demand for goods and services, and sales of tobacco \cite{olshansky1997ever,prescott1922law,peabody1924growth}.

Alongside the many contexts wherein Gompertz' law has been observed, an almost equal number of efforts have been made to interpret and derive this law from various phenomenological principles \cite{bajzer1997mathematical}. This effort has often been accompanied by a phenomenological comparison of another popular growth model called logistic growth, or Verhulst growth, where ``the power to avoid destruction'' decreases linearly with abundance, as opposed to logarithmically as in the Gompertz model. In a study comparing the Gompertz and the logistic growth models, Petroni et al. (2020) \cite{petroni2020logistic} show that Gompertz growth can be interpreted as \textit{maximally coherent} growth based on long-range interaction or correlation, while logistic growth is minimally coherent with non-collaborating elementary entities. With this interpretation, one is lead to postulate that the Gompertz model is a result of an external stimulus with perceived long-range interactions while the logistic model is a result of an internal stimulus with local interaction. As such, the logistic model has been popular both for communicable disease models \cite{harko2014exact,kroger2020analytical,schlickeiser2021analytical,heng2020approximately} and for predator-prey theory \cite{berryman1992orgins}, where disease or injury spreads through pairwise interactions. 

A number of studies in the biological context have argued that Gompertz growth could be activated from an external factor which is ubiquitously and instantaneously introduced to the system, e.g. as a toxic agent or an environmental stressor to which the system gradually adapts \cite{neafsey1988gompertz,neafsey1989gompertz,thompson1990biphasic}. In the original case of mortality rates as a function of age, this interpretation would imply that time is the environmental ubiquitous stressor or some proxy for time endogenous to the state of the system. Other environmental stressors like chronic radiation exposure have been observed to shift the Gompertz curve, an observation also theoretically developed by Sacher (1956) (\cite{sacher1956statistical}). Sacher modeled the toxic stressor as a stochastic perturbation to the physiologic state of a population wherein death of an entity would occur beyond some perturbation threshold. The interpretation of Gompertz growth as the result of a ubiquitous stochastic perturbation was recently revived by De Lauro et al. (2014) \cite{de2014stochastic} who arrived at the same result without alluding to a threshold value beyond which death was inevitable, but rather by letting the stochastic perturbation be the source of growth.

An alternative derivation locates the origin of Gompertz growth in extreme-value statistics rather than in interaction. Shklovskii (2005) \cite{shklovskii2005simple} derived the Gompertz mortality law from the Poisson probability of zero immune encounters, $n$, as $P(n=0)=e^{-N_0}$, with the mean rate $N_0(t)$ decaying linearly with age. Because death in this context can be viewed as a first-passage event beyond some threshold, this mechanism produces Gompertz dynamics without any interaction between entities. As we discuss below, the normalized Gompertz curve is identical to the cumulative distribution of a Gumbel (extreme-value) variable, so any process in which conversion time is the extreme of many independent sub-processes yields Gompertz growth. This provides a path to the Gompertz curve that does not rely on interaction, but rather on a common cause or stressor.

Another framework used to show how the Gompertz model can result from an external or ubiquitous stressor is chemical reaction theory. In reaction theory, a set of reactants combine to produce products, possibly in the presence of a catalyst. Markov (2019) \cite{markov2019reaction} showed that the Gompertz model can be interpreted as growth under a catalyst whose effect diminishes at a rate independent of the growth rate of the reactant. In other words, the catalyst can be considered as an external and ubiquitous stimulant or stressor to the reactants, perfectly in line with the ubiquitous stochastic perturbation model from Sacher and De Lauro et al. On the other hand, if the logistic growth equation is recast into the reaction network equivalent, the catalyst is diminished by the growth of the reactant, which suggests that the source of growth is internal or at least dependent on the internal state. So also in reaction theory the Gompertz model could be seen as a result of a ubiquitous field while the logistic model could be seen as a result of internal interactions of the reactants. 

The notion that Gompertz growth is maximally coherent and non-local has also been derived from quantum mechanical principles \cite{molski2003coherent}. Molski and Konarski (2003) showed that the Gompertz equation is the solution of a form of the time-independent Schrodinger equation. From this result, they derive the quantum coherent states which are non-local in space while moving along a classical time trajectory. This adds to our exhibits that the Gompertz model can be interpreted as non-local in origin. Coherent quantum states was originally formulated in the context of electromagnetic fields by \citet{glauber1963coherent} (for which he received the Nobel Prize). Some biological macroscopic growth phenomena are also due to such fields. For example, the role of internal electrical fields in regenerative growth is well established \citep{levin2021bioelectric} and is based on electrical potentials that act on sensitive cells \cite{becker1984electromagnetic}. In fact, it has been shown that cancer malignancies are mostly \textit{negative} in electric polarity with respect to the rest of the body, while nonmalignant (or benign) tumor pathology is mostly \textit{positive} in polarity \cite{langman1949technique}. This effect is also called \emph{depolarization} of the tumor with respect to surrounding tissue \citep{chernet2013transmembrane,cervera2020community}. Gompertz growth has also been widely observed in cancerous cells \cite{kozusko2003combining,laird1964dynamics}. One is then led to the interpretation that certain biological growth processes that exhibit Gompertz growth possibly are coupled with an electromagnetic field of specific character, or e.g. a collective nutrient deprivation. Consequently and unsurprisingly, the artificial disruption of such a field has been found to disrupt the mitotic cell cycle \cite{goodman1983pulsing,liboff1981alternating,norton1984bioelectric,liboff1984time}. 

This theory would also validate earlier cell-based theories where Gompertz growth in tumors are postulated to arise from an inverse relationship between diversity in cell states and the number of intercellular interactions \cite{kendal1985gompertzian}. In other words, where a maximally interacting system leads to a minimum of possible states the Gompertz emerges.

On the topic of disease modelling, the Gompertz model has been shown by Wang et al. \cite{wang2012richards} to be incompatible with a subset of models in the communicable disease paradigm. They used the common Susceptible-Infected-Recovered (SIR) model by \citet{kermack1927contribution} augmented by the Richards parameter \cite{richards1959flexible} and showed that the allowed parameter range within the this model does not cover the parameter value taken in the Gompertz scenario. Other work incorporates network heterogeneity, super-spreading, and behavioral feedback to model sub-exponential/Gompertz-like dynamics in transmission models \cite{vardavas2021modeling,osi2024parameter,lloyd2005superspreading}. If one incorporates a social proximity network on which the disease spreads, \citet{zonta2022virus} observed through computer simulations that a near-Gompertz growth can occur after an initial transient period but only for scale-free networks, a type of network that some claim to be rarely representative of social networks \cite{clauset2009power,broido2019scale,holme2019rare} and others claim to be common \citep{herrmann2020covid}, with varying degrees of rigor. We will show that one can obtain an asymptotically exact Gompertz curve from a ubiquitous stressor scenario based on microscopic interactions regardless of network topology, with deviations from exactness controlled by network heterogeneity and vanishing as the system synchronizes.

While many studies show the SIR model (and its extensions) as the fitted model in epidemics \cite{carletti2020covid,cooper2020sir,postnikov2020estimation,munoz2021sir,cooper2022dynamical,saikia2021covid}, some recent studies have successfully fitted a Gompertz model \cite{Ohnishi2020,Rypdal2020,Catala2020,rodrigues2020monte,Levitt2020}. 

\subsection{Main aim}
In this paper, our primary aim is to investigate the source of growth in self-organized processes and develop a microscopic version of the Richards model. In particular, we posit that the source of growth can be elucidated by considering interactions in the microscopic domain which can be aggregated to recover the traditional macroscopic growth equations. Second, we will elaborate on the idea that the logistic model results from local internal interactions while the Gompertz model results from a possibly external field or equivalently from an infinite set of higher order interactions between the microscopic entities involved in the growth. We will extend the Richards parametrization of the Gompertz-Logistic equations to apply in the microscopic domain with interacting entities, building on previous work in the macroscopic domain \citep{tjorve2017use,wang2012richards}. While the Gompertz has both been framed as a model with maximal coherence macroscopically through a resource-availability argument \citep{petroni2020logistic} and as a quantum-coherence reading of the same functional form \citep{molski2003coherent}, no known work derives a microscopic closure condition, a spectral gap of the decay modes, a mean-variance gap at $O(\theta)$, or a closure-error diagnostic, all of which are the quantitative, network-testable content of the present work. Thus, we present a novel bridge between the logistic and the Gompertz model which offers both a mathematical framework and a biological interpretation for the source of growth.

\subsection{Roadmap}
Our paper is structured as follows: we will first introduce the macroscopic Gompertz-Logistic models with the traditional Richards parameter as the bridge between them. We then present a microscopic extension adding a local coupling term between microscopic neighbors. The interpretation of such a coupling depends on which model we use. In the logistic extreme, the coupling term represents local/pairwise transmission propagation, as needed in e.g. contagion dynamics, while in a Gompertz extreme the same term represents the shared response to a common exposure where every entity is pulled toward the collective state equally regardless of which is ahead or behind, as if the stressor acts on every entity simultaneously rather than propagating from entity to entity.

We further show that the Gompertz model is 1) closed under microscopic aggregation, where the network's internal contributions vanish from macroscopic observables exactly for doubly stochastic coupling and asymptotically otherwise as the system synchronizes and 2) is consistent with a ubiquitous field or in the biological context: an environmental stressor external to the entities involved (animals, humans, cells, etc.). Then, we explore mathematical properties exhibited by the Gompertz model and prohibited by the logistic model, like stability in the face of internal fluctuations, mean-variance decoupling, both while introducing a microscopic interaction term in the model. We proceed with a discussion that summarizes our findings in the context of the other well-known growth modalities involving the Gaussian and the Laplacian model.

\section{Theory: From Macroscopic to Microscopic Richards growth}
\subsection{Macroscopic model}
\label{sec:macro}

First consider a system that grows according to logistic growth. If we let the variable $X(t)$ denote the abundance of a quantity in a normalized system where its maximum size is 1, the equation that determines its size as a function of time is
\begin{equation}
  \label{eq:logistic}
  \frac{1}{X}\frac{d}{dt} X(t) = \beta (1 - X(t)),
\end{equation}
where $\beta>0$ is a growth parameter. Here, the right hand side is commonly interpreted as representing ``resources available for growth'' which decrease linearly with respect to the relative abundance's growth rate on the left hand side. One way to generalize logistic growth is to introduce a parameter that exponentiates the resource availability in the following way,
\begin{equation}
\label{eq:Rich}
  \frac{1}{X}\frac{d}{dt} X(t) = \frac{\beta}{\theta} (1 - X^{\theta}(t)),
\end{equation}
also called $\theta$-logistic growth or Richards model \citep{richards1959flexible}. Petroni et al. (2020) argued that $\theta$ can be interpreted as a parameter capturing non-linear interaction effects. They showed that in the limit where the system becomes \textit{maximally cooperative} or \textit{coherent}, $\theta\rightarrow 0$ and the $\theta$-logistic equation reduces in this limit to the Gompertz growth model,
 \begin{equation}
 \label{eq:Gomp}
  \frac{1}{X}\frac{d}{dt} X(t) = \frac{d}{dt}\ln{X}(t) = - \beta \ln X(t).
\end{equation}
In this limit, the relative growth rate goes to infinity, $\lim_{X(t)\to 0^{+}}\frac{\dot X}{X} = \infty$, which could be seen as a verification of the maximally coherent characteristic of the system. One is reminded of a system that is perturbed and left to reach a new equilibrium. Petroni's definition of maximal coherence in the context of the Gompertz equation was shown to rigorously correspond to the definition of coherence used in quantum mechanical systems \cite{molski2003coherent} attributed to Glauber (1963) \cite{glauber1963coherent}.

It will also be useful to note that the Gompertz curve is an extreme-value distribution. 

Let $\xi_1,\dots,\xi_n$ be independent random times drawn from a distribution in the Gumbel domain of attraction, which includes the exponential, normal, and gamma, and let $M_n = \max_i \xi_i$. The Fisher–Tippett–Gnedenko theorem \citep{fisher1928limiting,gnedenko1943distribution} gives centering and scaling sequences $a_n, b_n$ such that

\begin{equation}
P\!\left(\frac{M_n - a_n}{b_n} \le t\right) \;\xrightarrow[n\to\infty]{}\; \exp\!\left(-e^{-t}\right),
\end{equation}
i.e. the Gumbel distribution. Writing this in unstandardized form with location $T$ and scale $\sigma$,

\begin{equation}
\label{eq:gumbel}
F(t) = \exp\!\left(-e^{-(t-T)/\sigma}\right),
\end{equation}
and comparing with the solution of Eq. \ref{eq:Gomp}, we see that $F(t) = X(t)/X_\infty$ identically, with $\sigma = 1/\beta$. Thus, the normalized Gompertz growth curve is the Gumbel cumulative distribution function. Equivalently, Gompertz growth is the law obtained when each entity's conversion time is the extreme of many independent sub-processes. Applying the double-log transform $g(F) = \ln(-\ln F)$ recovers the familiar diagnostic,

\begin{equation}
  g\big(F(t)\big) = -\beta\,(t - T),  
  \label{eq:linear-trans}
\end{equation}  
exactly linear in $t$ with slope $-\beta$.

The location parameter $T=\sigma \ln(n)$ in the exponential tail case, where $n$ is the number of sub-processes or barriers per entity. Increase in number of sub-processes therefore delays onset only logarithmically.
  
The Gompertz/Gumbel identity is important in our context because it involves no interaction between microscopic entities, but is rather a result of an extreme value statistics formulation. We used the maximum operator so the interpretation would be the "time at which the last barrier falls" for each entity. To aggregate to the population level, where each entity's conversion time is $\tau_i$ and following a Gumbel law, we just look at the empirical CDF which converges to the theoretical Gumbel CDF/Gompertz curve as N grows:
\begin{equation}
  \frac{X(t)}{X_\infty} \;=\; \frac{1}{N}\sum_i \mathbf{1}[\tau_i \le t] \;\xrightarrow[N\to\infty]{}\; P(\tau \le t) = F(t).
\end{equation} 
Notably, the macroscopic curve is deterministic in the large-N limit despite every entity being stochastic and no entity interacting with any other. However, convergence towards a Gumbel does require common scale and location parameters, whereby any discrepancy thereof will lead to non-linear observations after the double-log diagnostic in \ref{eq:linear-trans}. 

The dual construction using the minimum operator yields the classical Gompertz mortality law rather than the growth curve, $F(t) = 1 - \exp\!\left(-e^{(t-T)/\sigma}\right)$ and exponentially rising hazard. Here the relevant condition is on the left tail of the sub-process distribution due to the minimum operator, which must vanish faster than any polynomial. This is the structure underlying \citet{shklovskii2005simple}, where mortality is the first escape past immune clearance and the Poisson zero-encounter probability $P(0)=e^{-N_0(t)}$, with $N_0$ decaying linearly in age, supplies the required tail behavior.

Thus, the Gompertz process admits both a maximally correlated interpretation as seen in \citet{kendal1985gompertzian} and \citet{petroni2020gompertz} but also a completely uncorrelated interpretation as seen here. We shall see below that, even under a network coupling term, both these endpoints are indistinguishable in the aggregate, since at $\theta\to0$ the network's net contribution to aggregate growth vanishes identically (Eq. \ref{eq:redistributive}).

\subsection{Microscopic model}
Given that both the logistic and Gompertz models describe a sigmoidal saturation at the macroscopic level, what distinguishes them at the microscopic level? The answer will turn out to be the domain in which dynamics are linear, which in turn determines the correlation structure. First, note that the Gompertz function is naturally expressed as a linear differential equation in the log-domain $\ln{X(t)}$, closed under affine transformations, while the logistic function is naturally expressed as a polynomial in the original domain, not closed under such transformations. 

In the log-domain, the time derivative of the log-abundance, $\frac{d}{dt}\ln(X)=\frac{\dot{X}}{X}$, is the relative growth rate, unitless and automatically normalized. Thus, the Gompertz model is the only model in this family that is both linear and expressed in the natural unitless coordinate space. Linearity is a consequence of the log-transform's ability to transform multiplicative processes, that are common in biological growth phenomena, to additive processes. Any other value of $\theta$ in the Richards family, might be transformed to linearity but loses its anchor in the observable space. 

Let us split the system into $N$ microscopic abundance variables, $x_1(t), x_2(t), ..., x_N(t)$ (sometimes we omit $t$ for brevity), normalized to take values between 0 and 1, where 0 represents no realized growth while 1 represents the maximum potential growth having been realized. In other words, these variables represent the \textit{realized potential}, either in the form of individual volume, mass, likelihood of infection or mortality by being preyed on, etc. 

A key point is that Gompertz growth is \emph{closed} under the geometric mean of the microscopic equations in the natural (unitless) log-domain, i.e. both micro- and macroscopic domains obey the same dynamics. This is a natural consequence of the linearity of the log-coordinate system. To see this, note that macroscopic Gompertzian growth is retained when we compute the arithmetic mean in the log-domain, with $z_i = \ln x_i$, 
\begin{align}
\ln X(t) =& \frac{1}{N}\sum_i^N z_i(t)\\
\frac{d}{dt}z_i(t) =&-\beta z_i(t)\\
\therefore\frac{d}{dt}\ln X(t) = \frac{1}{N}\sum_i\frac{d}{dt}z_i(t) =& -\beta \frac{1}{N}\sum_iz_i(t)\\
\frac{d}{dt}\frac{1}{N}\sum_iz_i(t) =& -\beta \ln X(t).
\end{align}

The same closure does not hold for logistic dynamics. To see this, we must apply the microscopic logistic model with communication between $x_j$ and $x_i$ via a coupling constant $w_{ij}$ that is assumed both row- and column-normalized  $\sum_i w_{ij}=\sum_j w_{ij}=1$, i.e. the total influence received from node $j$ and returned out to the network across all nodes sums to 1,
\begin{align}
  \label{eq:microLogisticEpidemic}
  \dot{x_i} = \beta(1 - x_i)\sum_j w_{ij}x_j,
\end{align} 
which we name \emph{the microscopic transmission logistic model}.
This coupling matrix, $w_{ij}$ is doubly-stochastic \cite{sinkhorn1967concerning}, and it can be interpreted as a regular network, where each node shares the degree distribution, as opposed to a non-regular network where nodes have different degree distributions and strengths, although a non-regular network can be transformed to a doubly-stochastic network through Sinkhorn balancing. While we present most simulations with non-regular networks, we use the regular network assumption in the theoretical discussion as an idealized starting point. As shown in the sequel, only closure under aggregation depends on the doubly-stochastic assumption.

Notwithstanding network regularity, the coupling mechanism is absolutely necessary in a transmission model, but not in an external stressor model where a ubiquitous field can be regarded as the catalyst or trigger. So while we must include coupling with a network in a logistic model involving microscopic dynamics it is not necessary in Gompertzian dynamics. However, we shall see in the sequel that even when we include network coupling in the Gompertz model, closure still holds for some networks. 

Now, we proceed with the logistic model to average over the microscopic entities, with $\bar{x}=\frac{1}{N}\sum_ix_i$,
\begin{equation}
  \label{eq:microDeterministic}
  \frac{1}{N}\sum_i\dot{x_i} = \frac{\beta}{N}\sum_i (1-x_i)\sum_j w_{ij} x_j = \frac{\beta}{N}\sum_i\sum_j w_{ij} x_j - \frac{\beta}{N}\sum_i\sum_j w_{ij} x_i x_j.
\end{equation}
The first term is simply $\frac{\beta}{N}\sum_j x_j = \beta\bar{x}$.

Now, for closure to hold, we need the right hand side to equal $\beta \bar{x}(1 - \bar{x})$. The second term is the weighted average of pairwise products, $\frac{\beta}{N}\sum_i\sum_j w_{ij} x_i x_j$, which is $\beta \bar{x}^2$ only if either $x_i$ and $x_j$ are uncorrelated across the network, or all nodes are completely synchronized i.e. $x_i=\bar{x}$ for all $i$. Note that non-regular networks also exhibit this lack of closure in the logistic setting.

Neither synchronicity nor lack of correlation is compatible with most scenarios the microscopic logistic model is meant for, e.g. predator-prey dynamics or communicable diseases. Synchronization precludes any heterogeneity in abundance or health status, while the uncorrelated condition requires that abundance be uncorrelated with its neighbors — contradicting the very transmission mechanism on which communicable disease models are predicated and similarly for predator-prey models. In a real epidemic, highly connected nodes are infected earlier, creating strong positive network-state correlations that violate both conditions. Previous work has shown near-Gompertz growth with a scale-free network in computer simulations \cite{zonta2022virus}, but theoretical equivalence has yet to be shown and transient non-Gompertz behavior is seen initially in those simulations. Scale-free networks are also contested in social networks \cite{clauset2009power,broido2019scale,holme2019rare}.

Instead, the proposed microscopic Gompertz model implies higher-order dependence between microscopic variables without any new parameters. This is seen by examining the partial derivatives of the geometric mean with respect to the abundance variables,
\begin{align}
X(t) =& \left[\prod_i x_i(t)\right]^{1/N}\\
\therefore \frac{\partial}{\partial x_i}X(t) =& \frac{1}{N}\frac{X}{x_i},
\end{align}
and so any subsequent partial derivative with respect to another $x_j \neq x_i$ will also be a function of all $(x_1, x_2, \ldots)$, with no order at which the dependence truncates. This is a mathematical property of the geometric mean as an aggregation operator: the log-normal distribution in the abundance domain has nonzero cumulants at all orders, unlike the Gaussian distribution that arises from arithmetic averaging. In contrast, the arithmetic mean has $\partial X/\partial x_i = 1/N$ with all higher mixed partials equal to zero, so the logistic model's macroscopic observable couples to microscopic states only at first order.

We emphasize that this infinite-order dependence is a necessary consequence of the dynamics being linear in the log-domain. The microscopic Gompertz dynamics in $z_i = \ln x_i$ are linear and only pairwise-coupled, producing a jointly Gaussian distribution in $z$-space with correlations fully characterized at second order. The exponential map $x_i = e^{z_i}$ then transforms this second-order structure into the infinite-order cumulant structure of the log-normal distribution in the original domain. Thus, "infinite correlations" are generated by the coordinate transformation between the natural dynamical domain ($z$-space) and the observable domain ($x$-space), not by explicit higher-order interaction terms in the equations of motion.

For the microscopic Gompertz model, the aggregation operator is determined by the condition that the dynamics should be closed under aggregation in accordance with observations, thus the geometric mean in the natural choice. That logistic growth does not close exactly at the macroscopic level implies some noise is expected in real observations or simulations (see Results section). 

In brief, the emergence of the Gompertz curve at the macroscopic or collective level suggests that the system is coherent, accordance with previous work \cite{zhang1994log,petroni2020gompertz,molski2003coherent}, presumably as a result of the simultaneous exposure to the same underlying stressor and due to the implied log-normal nature of the microscopic entities, where multiplication replaces addition as the aggregating operator. Thus, Gompertzian growth can either be seen as infinitely correlated growth as a natural consequence of the dynamics, or as independent growth through an underlying common stressor as seen with the Gumbel connection. We have thus showed \emph{what} distinguishes the logistic from the Gompertz model, namely the closure conditions and the correlation order, and now turn to show \emph{how} these models connect through the continuous Richards parameter.

\section{A microscopic Richards family and its Gompertz limit.}

The Richards model unifies the macroscopic logistic and Gompertzian growth models, but by proposing a microscopic version of this model we shall see that Gompertz growth at the microscopic level is the only type of growth which is coherent in the sense that a) internal fluctuations do not perturb the macroscopic mean, b) the macroscopic mean is decoupled from the variance of the fluctuations, and c) no entity can dominate the collective forcing, and the network's net contribution to aggregate growth vanishes exactly for doubly stochastic coupling, and asymptotically on general networks as the system synchronizes. Let us start with N nodes in the log-transformed domain, $z_i = \ln x_i$, connected by a network of edges represented by our doubly stochastic connectivity matrix, $W$, where the $ij^{th}$ element is the positive coupling strength, between node $i$ and $j$. With these assumptions, the following microscopic model for growth is postulated,
\begin{equation}
  \label{eq:newMicro}
  dz_i = \left[\frac{\beta}{\theta}(1 - e^{\theta z_i}) + \beta(\bar{z}_{\theta,i} - z_i)\right]dt + \sigma\,dB_i(t)
\end{equation}
where $\bar{z}_{\theta,i} = \frac{1}{\theta}\ln(\sum_j w_{ij} e^{\theta z_j})$ is the order-$\theta$ power mean of neighbors' states. The parameter $\theta$ smoothly interpolates between Gompertz ($\theta \to 0$) and logistic ($\theta = 1$) dynamics, with intermediate values recovering the Richards growth family. 

For $\theta > 0$, the coupling term depends on the identity and state of the node's particular neighbors, and is dominated by those furthest ahead. At $\theta\rightarrow0$, while the coupling term is felt by every individual node, it contributes exactly zero to the aggregate growth (shown below). The Gompertz limit is therefore the unique member of the family in which the transmission network becomes dynamically invisible in the aggregate when using the natural unitless log-coordinate. We stress that this coupling term is the attraction to the collective state itself, $\bar{z}_{\theta,i}$, not an exogenous forcing term; the model itself does not impose a stressor but rather shows how a system might react to one -- a reaction which is non-local in nature. And the same Gompertz curve arises from no coupling at all as shown with the Gumbel distribution above, so both coupled and uncoupled constructions are indistinguishable in the aggregate.

We note that the $\theta = 1$ endpoint of Eq. \ref{eq:newMicro} is not the transmission logistic model of Eq. \ref{eq:microLogisticEpidemic}. Working out the endpoint of Eq. \ref{eq:newMicro} gives $\dot{x}_i = \beta x_i(1-x_i) + \beta x_i\ln(\sum_j w_{ij}x_j/x_i)$, which combines intrinsic logistic growth with a multiplicative synchronizing coupling. This differs structurally from Eq. \ref{eq:microLogisticEpidemic}, where growth is driven entirely by transmission from neighbors and there is no intrinsic self-term. The interpolating family of Eq. \ref{eq:newMicro} is thus a family of coherent growth models — intrinsic dynamics synchronized by coupling — parameterized by the interaction order $\theta$. The transmission logistic of Eq. \ref{eq:microLogisticEpidemic} lies outside this family: it is the model driven solely by transmission lacking an intrinsic growth term, and as shown above it does not close under aggregation.

The coupling is designed so that $\theta$-dependence is consistent with the macroscopic Richards interpolation. The noise is herein assumed to follow standard Wiener increments, and its structure is multiplicative in the original domain and thus retains positive abundance values. Note that for perfectly synchronized nodes, with $z_i = \bar{z}, \;\forall i$, Equation \ref{eq:newMicro} reduces to a single-entity model, as seen in \citet{de2014stochastic}.

In the Gompertz limit, i.e. $\theta \rightarrow 0$, we see first that the model is still closed under aggregation even with the coupling term. Upon averaging over all nodes, the coupling term, $\beta(\bar{z}_{\theta,i} - z_i)dt$, vanishes because the doubly stochastic property gives $\frac{1}{N}\sum_i\sum_jw_{ij}z_j=\bar{z}$. We can then write the average over $z_i$ in Eq. \ref{eq:newMicro} as
\begin{equation}
  \label{eq:newMicro2}
  d\bar{z} = -\beta \bar{z}dt + \frac{\sigma}{N}\sum_i dB_i(t),
\end{equation}
which is a classical 1D Ornstein–Uhlenbeck process with variance $\frac{\sigma^2}{2\beta N}$, where the independent noise term sum, $\frac{1}{N}\sum_i dB_i$, has quadratic variation $\frac{1}{N}dt$ and is therefore equal in law to $\frac{1}{\sqrt{N}}d\tilde{B}$ for a single Wiener process $\tilde{B}$. Thus $d\bar{z}$ is governed by a one-dimensional OU process with effective noise amplitude $\sigma/\sqrt{N}$.

Secondly, any microscopic fluctuations revert to the mean, producing a stable evolution independent of the fluctuation profile. We will show this assuming deterministic regime ($\sigma=0$), but the analysis can be extended to the stochastic regime. First, define a set of zero-mean fluctuations 
\begin{equation}
  \delta_i = z_i - \bar{z}.
\end{equation}
then, with $\bar{z}_{\theta,i}\xrightarrow[\theta \to 0]{} \sum_j w_{ij}z_j$, and $d\bar{z}/dt\xrightarrow[\theta \to 0]{} -\beta\bar{z}$,
\begin{align}
\frac{d\delta_i}{dt} &= \frac{dz_i}{dt} - \frac{d\bar{z}}{dt} \\
&= -\beta z_i + \beta\left(\sum_j w_{ij}z_j - z_i\right) + \beta\bar{z} \\
&= -2\beta(\bar{z} + \delta_i) + \beta\sum_j w_{ij}(\bar{z} + \delta_j) + \beta\bar{z} \\
&= -2\beta\bar{z} - 2\beta\delta_i + \beta\bar{z}\sum_j w_{ij} + \beta\sum_j w_{ij}\delta_j + \beta\bar{z} \\
&= -2\beta\delta_i + \beta\sum_j w_{ij}\delta_j,
\end{align}
where we have used the assumption that rows of $w_{ij}$ sum to 1 and $\sigma=0$. 
The result is a linear system $\dot{\boldsymbol{\delta}} = \beta(-2I + W)\boldsymbol{\delta}$. For a regular network, $W$ is symmetric and so the eigenvalues are real and lie in $[-1, 1]$, so the eigenvalues of $-2I + W$ lie in $[-3, -1]$. All eigenvalues are strictly negative, meaning every deviation mode decays exponentially with rate at least $\beta$. Thus, one could say the system is unconditionally stable, where deviations shrink regardless of their initial magnitude. For the non-regular, row-normalized couplings used in the simulations, W is not normal and may have transient instability due to non-orthogonal eigenvectors, but eigenvalues will still be real and in $[-1,1]$ due to the similarity between the asymmetric weight matrix and it's symmetric counterpart, i.e. $D^{-1}A$ is similar to $D^{-1/2}AD^{-1/2}$. Simulations show that transient effects are small.

Combining the first and the second point we see that the Gompertz model does not require synchronicity, but rather imposes it. One could call this a characteristic of a coherent system. 

Thirdly, the Gompertz model's macroscopic mean and microscopic variance are decoupled, a property that is absent for any other model in the Richards spectrum where $\theta > 0$, in the natural log-coordinate. To demonstrate this, let $\bar{z} = \frac{1}{N}\sum_i z_i$ define the global mean field and $\delta_i = z_i - \bar{z}$ denote the zero-mean microscopic fluctuations. We define the localized, network-weighted variance of node $i$ as $s_i^2 \equiv \sum_j w_{ij}\delta_j^2$. Since $W$ is a row-normalized matrix, its rows sum to unity ($\sum_j w_{ij} = 1$). Expanding the microscopic order-$\theta$ mean around the Gompertz limit ($\theta \to 0$) yields:

\begin{align}
\bar{z}_{\theta,i} &= \frac{1}{\theta}\ln\left(\sum_j w_{ij} e^{\theta (\bar{z} + \delta_j)}\right) = \bar{z} + \frac{1}{\theta}\ln\left(\sum_j w_{ij} e^{\theta \delta_j}\right) \\
&= \bar{z} + \frac{1}{\theta}\ln\left( \sum_j w_{ij} \left[1 + \theta \delta_j + \frac{\theta^2}{2} \delta_j^2 + O(\theta^3)\right]\right) \\
&= \bar{z} + \frac{1}{\theta}\ln\left( 1 + \theta \sum_j w_{ij}\delta_j + \frac{\theta^2}{2} s_i^2 + O(\theta^3)\right).
\end{align}
Utilizing the standard logarithmic expansion $\ln(1 + x) = x - \frac{1}{2}x^2 + O(x^3)$, we obtain:
\begin{equation}
\bar{z}_{\theta,i} = \bar{z} + \sum_j w_{ij}\delta_j + \theta \left[ \frac{1}{2}s_i^2 - \frac{1}{2}\left(\sum_j w_{ij}\delta_j\right)^2 \right] + O(\theta^2). \label{eq:meanVarianceCoupling}
\end{equation}

Crucially, Eq. (\ref{eq:meanVarianceCoupling}) illustrates that the deterministic macroscopic drift decouples from the microscopic variance terms, $s_i^2$, only in the exact Gompertz limit where $\theta \to 0$. For any finite $\theta > 0$, the macroscopic dynamics remain inherently coupled to the higher-order moments of the microscopic state distribution.

Consequently, for the general Richards growth framework, the microscopic equations reduce strictly to the macroscopic phenomenological model generally only if the system exhibits complete phase synchronicity ($z_i = \bar{z}$, or equivalently, $\delta_i = 0$). While the macroscopic Richards model is strictly \emph{conditioned} on this synchronicity, the Gompertz model actively \emph{imposes} it in the deterministic case ($\sigma = 0$). 

In the stochastic regime the drift reduction remains exact, while the mean-field itself retains a residual fluctuation. Since the $B_i$ are independent, the aggregated noise has amplitude $\sigma/\sqrt N$, and $\bar z$ is an Ornstein–Uhlenbeck process with stationary variance $\mathrm{Var}(\bar z) = \sigma^2/(2\beta N)$, vanishing in the large-$N$ limit.

Lastly, we show that each entity equidistant from the mean contributes equally to the growth in the Gompertz model. To show this, we just have to show that this collective growth term, $\beta (\bar{z}_{\theta\rightarrow0,i} - z_i)$ is independent of $z_i \; \forall i$ and only a function of the fluctuations, $\delta_i$. We use the fluctuation from the mean $z_i = \bar{z} + \delta_i $,  viz.
\begin{align*}
  \bar{z}_{\theta\rightarrow0,i} - z_i &= \lim_{\theta\rightarrow0}\frac{1}{\theta}\ln\left(\sum_jw_{ij}e^{\theta z_j}\right) - z_i \\
  &= \sum_jw_{ij} z_j - z_i \\
  &= \sum_jw_{ij} (\bar{z} + \delta_j) - (\bar{z} + \delta_i) \\
  &= \sum_jw_{ij} \delta_j - \delta_i,  \\
\end{align*}
which completes the argument.

In contrast, logistic growth ($\theta=1$) has a coupling term $\bar{z}_{1,i}=\ln(\sum_j w_{ij} e^{z_j})$ which depends on all neighbours and any node $z_j \gg \bar{z}$ will dominate the exponential sum. Thus, in logistic growth the outliers drive the growth of the system, while in Gompertz growth, outliers only affect growth linearly. 

Moreover, while each node sees this network term, the aggregate does not. Averaging the collective forcing over all nodes and using column-normalization ($\sum_i w_{ij} = 1$) together with $\sum_i \delta_i = 0$,
\begin{equation}
\frac{1}{N}\sum_i \left(\sum_j w_{ij} \delta_j - \delta_i\right)
= \frac{1}{N}\sum_j \delta_j \sum_i w_{ij} - \frac{1}{N}\sum_i \delta_i = 0.
\label{eq:redistributive}
\end{equation}
 The network is therefore \emph{purely redistributive} at $\theta \to 0$: it exchanges growth between nodes but contributes nothing to the growth of the collective, which is governed entirely by each entity's own drift $-\beta z_i$.

This property is unique to the Gompertz limit when using the natural log coordinate system. For $\theta > 0$, Eq. (\ref{eq:meanVarianceCoupling}) shows that the collective forcing carries an additional term of order $\theta$,
\begin{equation}
\bar{z}_{\theta,i} - z_i = \left(\sum_j w_{ij}\delta_j - \delta_i\right) + \frac{\theta}{2}\left[s_i^2 - \left(\sum_j w_{ij}\delta_j\right)^2\right] + O(\theta^2),
\end{equation}
whose average over nodes does not vanish, since it depends on the network-weighted variances $s_i^2$ rather than on the zero-mean fluctuations alone. For any $\theta > 0$ the network thus injects growth into the aggregate through the microscopic variance, and the amount injected depends on the topology through the weights $w_{ij}$.

The aggregate growth law is therefore independent of the network: at $\theta \rightarrow 0$ the collective trajectory is generated entirely by the entity-level drift, and the coupling only redistributes among entities. This is consistent with the extreme-value construction earlier (Sect. \ref{sec:macro}), where the identical macroscopic curve arises with no coupling at all. It does not by itself exclude transmission, a conservative transport process can also be aggregate-neutral, but it does show that no network mechanism is required to produce the Gompertz curve.

\subsection{Dropping the assumption of double stochasticity}
 Without the doubly stochastic network assumption both the mean-variance decoupling and the mean reversion of the fluctuations still hold as they require only row-normalization. Only the first property, closure under aggregation, will retain the extra term $\frac{1}{N}\sum_i\beta (\bar{z}_{\theta,i} - z_i)$, but due to the synchronicity property for the Gompertz model, this term will go to zero as time progresses in the deterministic case and a stationary distribution in the stochastic case. The rate of this decay, or the closure discrepancy, depends on network heterogeneity in the following way. Letting $k_i$ be the number of neighbors for node $i$, $\mathcal{N}(i)$ be the set of those nodes, and $c_j=\sum_{i\in \mathcal{N}(j)}1/k_i=\sum_i w_{ij}$ be the network weight for this node, 
\begin{equation}
  \frac{\epsilon_G}{\beta} \approx \left|\frac{1}{N}\sum_i (\bar{z}_{\theta,i} - z_i)\right| = \frac{1}{N}\sum_i\left(\sum_j w_{ij}z_j - z_i\right) = \frac{1}{N}\sum_j z_j\underbrace{\sum_i w_{ij}}_{=\,c_j} - \frac{1}{N}\sum_i z_i = \frac{1}{N}\sum_j (c_j-1)z_j,
\end{equation}
where the random error terms $dB_i(t)$ average to zero approximately (hence the approximate-equality sign). And since $z_j$ can be decomposed into a mean and fluctuation term, $\bar{z}_{\theta,j} + \delta_j$, whereby the mean will cancel in the summation, we can express the discrepancy approximately as a function of the covariance between the degree heterogeneity and the node fluctuations 
\begin{equation}
  \epsilon_G \approx \beta \left|\frac{1}{N}\sum_j(c_j - 1)\delta_j\right| = \beta|\text{Cov}(\mathbf{c},\boldsymbol{\delta})|,
\end{equation}
since $\bar{\mathbf{c}}=1$, and with a slight abuse of notation as these are not random vectors, but rather statistical properties. The approximation will be exact in the deterministic case. Also, note that for a regular homogeneous network $c_j=1 \; \forall j$, resulting in zero closure error.

We now see that both large network heterogeneity or large node variance results in large deviations from the macroscopic model. For example, consider a network where there is a local group of nodes with low degrees and few neighboring nodes. Such low-degree nodes will synchronize amongst themselves but not as strongly towards the global mean, thus creating a higher covariance and closure discrepancy. However, simulation results show that the mean-field approximation is good even for quite heterogeneous, non-regular real networks.

The corresponding closure discrepancy for logistic growth depends directly on the node states' cross-covariance, shown from Eq. \ref{eq:microDeterministic}, 
\begin{equation}
  \epsilon_L \approx \beta\left|\frac{1}{N}\sum_j\sum_iw_{ij}x_ix_j - \bar{x}^2\right| = \beta|\text{Cov}_W(\mathbf{x},\mathbf{x})|,
\end{equation}
where the covariance includes the network weights for each node. The logistic model's suspected larger closure discrepancy arises from three reinforcing mechanisms — mean-variance coupling that prevents the macroscopic equation from closing, absence of the self-correcting synchronization that the Gompertz model uses to eliminate network-dependent coupling, and asymmetry across nodes exacerbated through the exponential aggregation, especially when there is a large growth velocity and microscopic variance with extreme values dominating. These three effects are absent in the Gompertz limit, which is why the closure discrepancy is both smaller and self-correcting.

 In brief, neither Gompertz nor logistic close without a mean-field approximation, i.e. for non-regular networks, although the logistic model exhibits more instability than the Gompertz in simulations (see Fig. \ref{fig:closure_error_comparison}). With regular networks, the Gompertz exhibits closure while the logistic model still enjoys no such property. Further elaboration on rate of decay of closure discrepancy is left for future work.

\section{Classifying the Gaussian and Exponential models with the Logistic and the Gompertz models}

Up until now we have seen how logistic growth emerges from pairwise correlated growth while Gompertz growth emerges from dynamics linear in the log-domain, aggregate-indistinguishable between correlated and independent entities. But what if there are other dependency structures? We can for example formulate a growth equation where the relative growth is only linearly dependent on time, where we make a slight modification and model the decay or \textit{potential abundance}, $s(t) = 1 - x(t)$, such that $s(0)=1$,
\begin{equation}
  \frac{d}{dt}s(t) = - \beta t s(t),
\end{equation}
or 
\begin{equation}
  \frac{d}{dt}\ln s(t) = - \beta t,
\end{equation}
for some constant $\beta>0$. 
 
Solving this equation yields the familiar Gaussian function, 
\begin{equation}
  s(t) = A e^{ - \beta t^2 / 2},
\end{equation}
where $\beta$ can be interpreted as the inverse variance or dispersion of the growth phenomena and where $A=1$ when $s(0)=1$. Thus, not surprisingly, Gaussian growth emerges for mutually independent events that have a linear time dependence. The same result was originally formulated by Gauss with space as the independent variable instead of time and with spatial estimation error as the dependent variable instead of abundance \cite{gauss1857theory} (see \cite{stahl2006evolution} for details)\footnote{Another example of space as the dependent variable in the Gompertz setting is given by \citet{molski2003coherent}.}

If we remove the dependency on time altogether, we observe the exponential growth equation
\begin{equation}
  \frac{d}{dt}\ln s(t) = - \beta,
\end{equation}
which yields 
\begin{equation}
  s(t) = A e^{ - \beta t}.
\end{equation}
This exponential equation implies independence of both time and the other entities. In the context of measurement error this function would represent the Laplacian error law which states that absolute measurement error decreases linearly with the error itself \footnote{Or as Laplace originally stated: ``[...] we have no reason to suppose a different law for the ordinates than for their differences'' \cite{stahl2006evolution}.}. 

To summarize we have
\begin{equation}
\frac{d}{dt}\ln s(t) = -\beta \cdot g(s, t)
\end{equation}
where $g$ encodes growth rate dependency, or the source of growth:

\begin{itemize}
\item Exponential: $g = 1$ (no dependence)
\item Gaussian: $g = t$ (time dependence only)
\item Logistic: $g = 1 - s$ (abundance dependence only, linear)
\item Gompertz: $g = - \ln s$ (abundance dependence only, logarithmic)
\end{itemize}

The Gompertz entry integrates to $s(t) = \exp\!\big(e^{\beta t}\ln s(0)\big) = \exp\!\big(-e^{\beta(t - T)}\big)$, with $T = -\beta^{-1}\ln(-\ln s(0))$, which is the survival function of the \emph{minimum}-operator Gumbel law and carries the exponentially rising hazard $-\frac{d}{dt}\ln s = \beta\, e^{\beta(t-T)}$ of Gompertz's original mortality law with $T$ acting as a de facto scaling parameter. This is the dual of the \emph{maximum}-operator Gumbel growth curve $X(t)/X_\infty = \exp\!\big(-e^{-\beta(t - T)}\big)$ of Section \ref{sec:macro}: the two constructions share the same double-log diagnostic up to the sign of the exponent, corresponding to whether the modeled quantity is decay ($s$, first barrier to fall) or growth ($X$, last barrier to fall).

Note also that closure under aggregation is only relevant where there is inter-entity coupling, and since neither the Exponential or Gaussian scenarios have such coupling, closure occurs trivially, and microscopic treatment thereof becomes superfluous.
\section{Simulations and Results}

\begin{figure}[H]
  \centering
  \includegraphics[width=15.4cm]{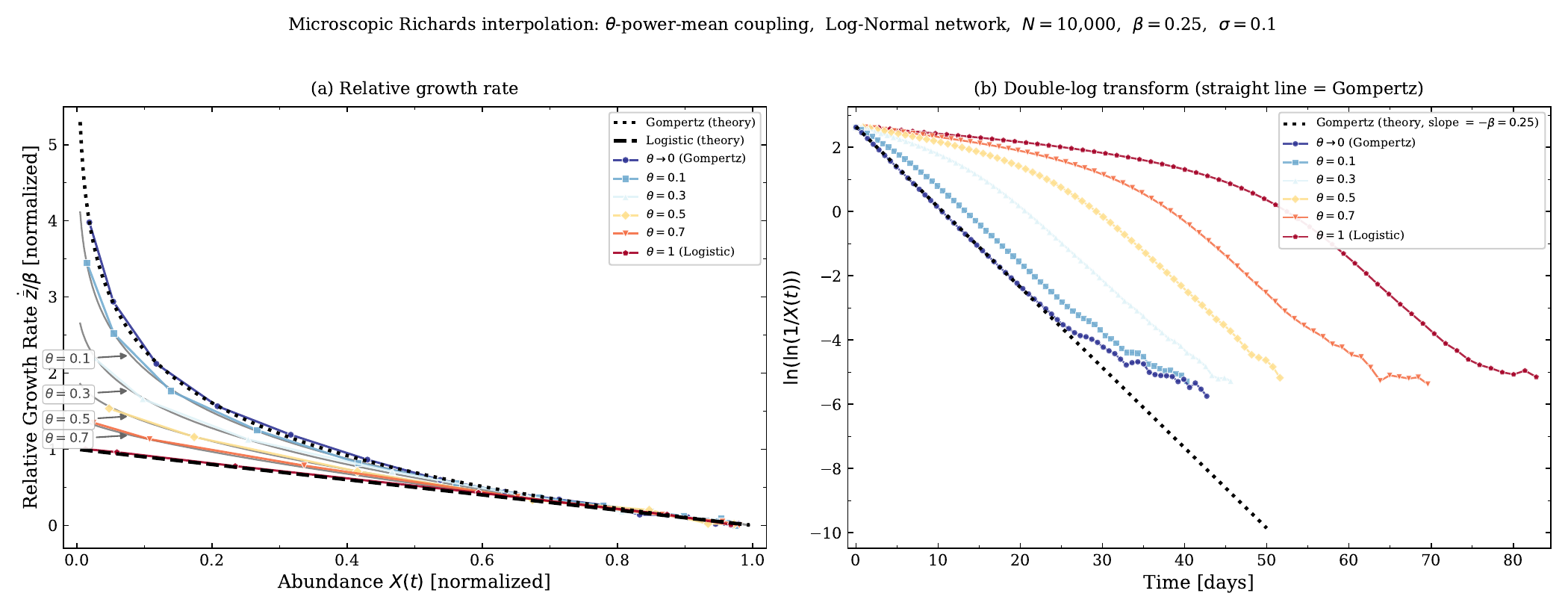}
  \caption{Different simulated Richard growth models are compared with their theoretical curves, using a Log-Normal network with 10,000 nodes and mean-field approximation. (a) Relative growth plotted against abundance and (b) Double-log transform $\ln(\ln(1/X(t)))$ versus time, where a straight line indicates Gompertz growth. Note that the higher values of $\theta$ makes the growth unstable due to higher mean-variance coupling as discussed in the text.}
  \label{fig:gompertz_richards_interpolation}
\end{figure}

Simulated microscopic growth scenarios for a variety of Richards "coherence" $\theta$ parameters all show close correspondence to their theoretical counterparts, except at high values of $\theta$ where the mean-variance coupling identified in Eq. \ref{eq:meanVarianceCoupling} produces visible deviations from theory, particularly in the late stages where the double-log transform amplifies residual fluctuations near saturation (Figure \ref{fig:gompertz_richards_interpolation}). A time-step convergence analysis confirms that these high-$\theta$ deviations are a genuine structural effect (the mean-variance coupling of Eq. \ref{eq:meanVarianceCoupling}) rather than Euler--Maruyama discretization error: as $\Delta t \to 0$ the Gompertz model ($\theta \to 0$) converges to zero deviation, whereas the $\theta > 0$ deviation persists at a nonzero floor that grows with $\theta$ (see Appendix).

\begin{figure}[H]
  \centering
  \includegraphics[width=15.4cm]{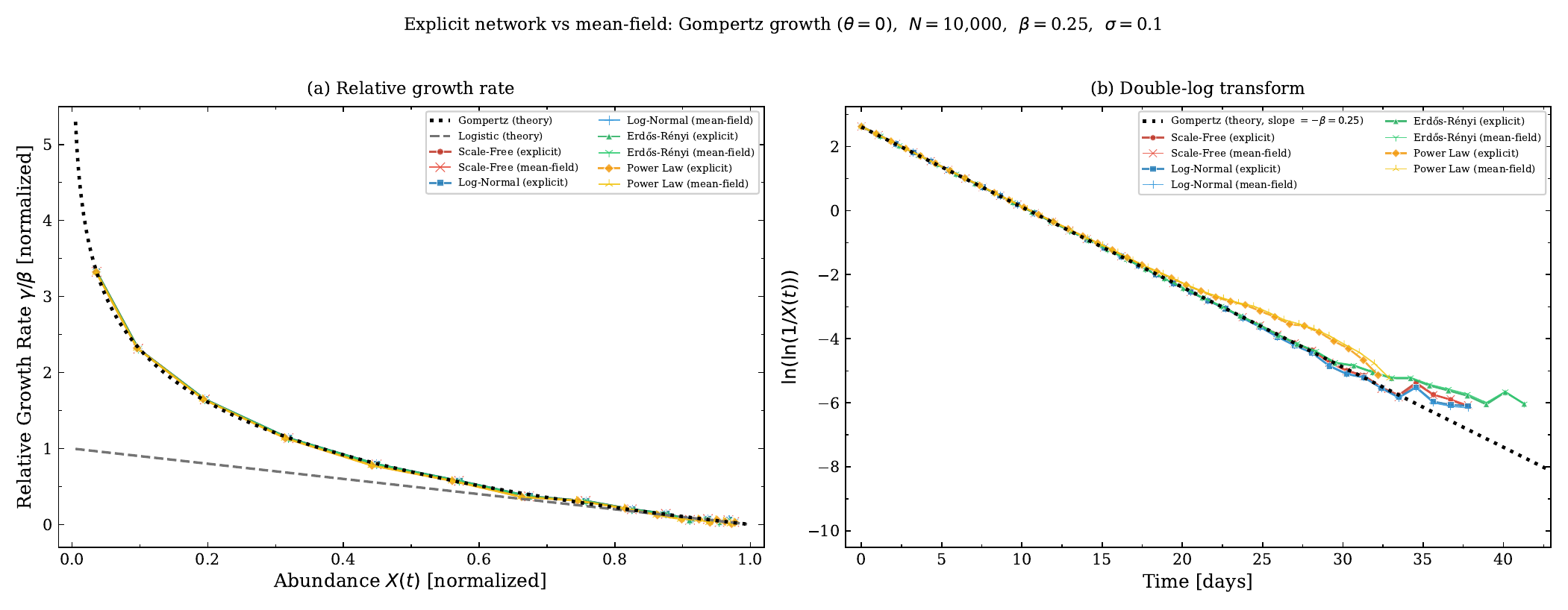}
  \caption{Comparison of Gompertz growth ($\theta = 0$) on explicit configuration-model networks versus the mean-field approximation, for four network topologies: Scale-Free ($P(k) \propto k^{-3}$), Log-Normal, Erdős-Rényi, and Power Law ($P(k) \propto k^{-2.2}$). Each network has $N = 10{,}000$ nodes with row-normalized coupling $w_{ij} = a_{ij}/k_i$, where $a_{ij}$ is the adjacency matrix and $k_i = \sum_j a_{ij}$ is the degree of node $i$. This normalization is not doubly stochastic for heterogeneous degree distributions. The explicit simulations compute each node's neighbor average from its actual neighbors in the sparse adjacency structure, with no global averaging. (a) Relative growth rate $\gamma/\beta$ versus normalized abundance $X(t)$. (b) Double-log transform $\ln(\ln(1/X(t)))$ versus time, where a straight line indicates Gompertz growth. Parameters: $\beta = 0.25$, $\sigma = 0.1$, $\Delta t = 0.02$.}
  \label{fig:gompertz_explicit_vs_meanfield}
\end{figure}

\begin{figure}[H]
  \centering
  \includegraphics[width=15.4cm]{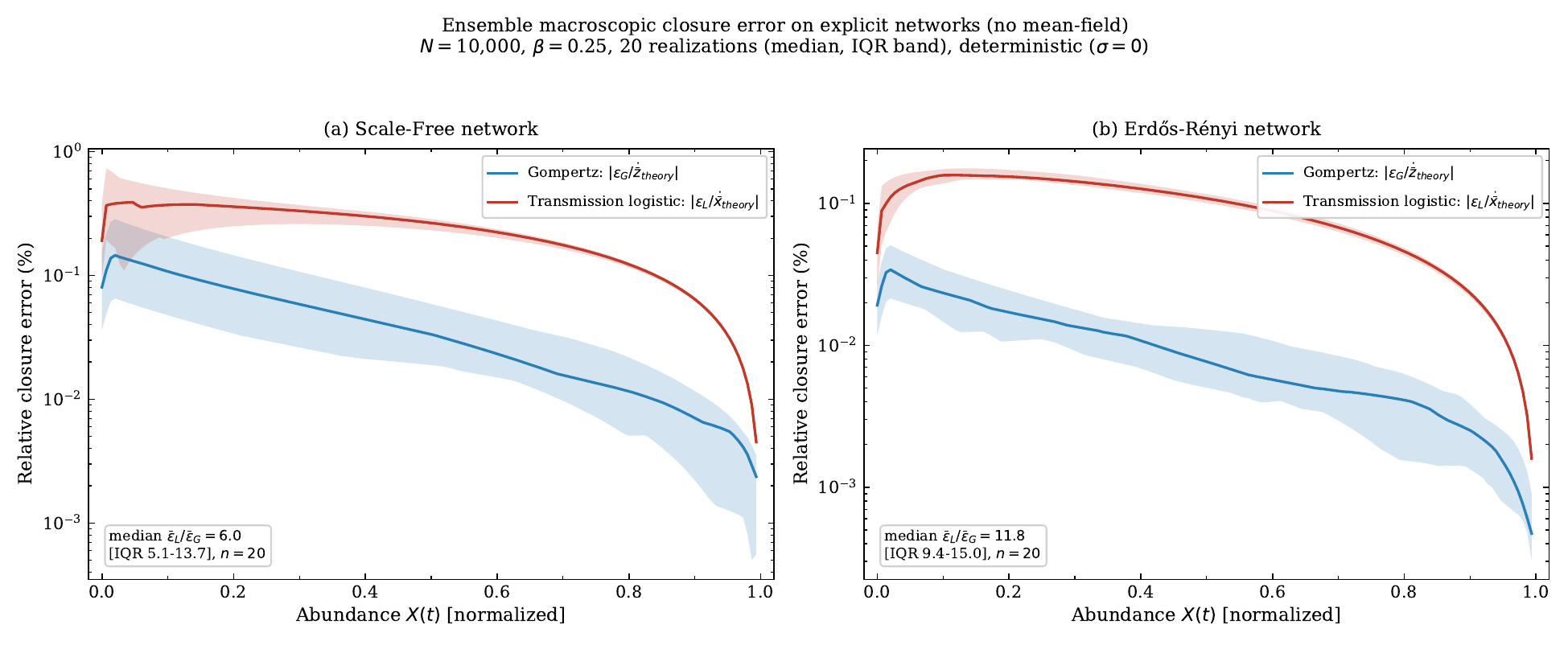}
  \caption{  
  Comparing relative macroscopic closure error for the canonical transmission logistic (Eq. \ref{eq:microLogisticEpidemic}) against the canonical Gompertz ($\theta = 0$ limit of Eq. \ref{eq:newMicro}) on non-regular networks without a mean-field assumption with row-normalized coupling $w_{ij} = a_{ij}/k_i$, where $a_{ij}$ is the adjacency matrix and $k_i = \sum_j a_{ij}$ is the degree of node $i$. The Gompertz closure error is defined as $|\epsilon_G / \dot{\bar{z}}_{\text{theory}}|$ where $\epsilon_G = \dot{\bar{z}} - (-\beta\bar{z})$ measures the deviation of the actual macroscopic dynamics from the closed-form Gompertz equation. The logistic closure error is $|\epsilon_L / \dot{\bar{x}}_{\text{theory}}|$ where $\epsilon_L = \dot{\bar{x}} - \beta\bar{x}(1-\bar{x})$ measures the deviation from the closed-form logistic equation (under the arithmetic mean $\bar{x}$). All are computed on configuration-model random graphs with $N = 10{,}000$ nodes on the deterministic (noise-free) dynamics, since closure under aggregation is a property of the deterministic vector field. Solid curves are the median over $20$ independent network and initial-condition realizations and the shaded bands span the interquartile range ($25$--$75$th percentiles); the two models share the same initial condition within each realization. (a) Scale-Free network ($P(k) \propto k^{-3}$, max degree $\sim 10^2$--$10^3$ across realizations). (b) Erdős-Rényi network ($k \sim \text{Poisson}(\lambda=6)$, max degree $\approx 15$--$19$). The annotated $\bar{\epsilon}_L/\bar{\epsilon}_G$ is the median, across realizations, of the ratio of the logistic model's relative closure error to the Gompertz value, averaged over the growth trajectory (the mean of $|\epsilon/\dot{X}_{\text{theory}}|$ over $0 < X < 1$), with the interquartile range in brackets. Parameters: $\beta = 0.25$, $\Delta t = 0.02$, $\sigma = 0$ (deterministic).}
  \label{fig:closure_error_comparison}
\end{figure}

Comparing the mean-field approximation with a non-approximating explicit network simulation does not yield notable discrepancies (Figure \ref{fig:gompertz_explicit_vs_meanfield}), further validating the approximation scheme. The explicit and mean-field results are indistinguishable across all topologies, including the Scale-Free network (max degree 417) and Power Law network (max degree 14{,}675), confirming that the macroscopic Gompertz dynamics are robust to actual network structure and only weakly sensitive to the doubly stochastic or mean-field assumptions (also shown in Figure \ref{fig:closure_error_comparison}).

The most informative closure comparison is between the stressor and communicable-disease paradigms. Thus, we want to compare the microscopic Gompertz model in Eq. \ref{eq:newMicro} with $\theta=0$ and the canonical transmission logistic model given in Eq. \ref{eq:microLogisticEpidemic} as opposed to the $\theta = 1$ limit of Eq. \ref{eq:newMicro}. This is because the canonical transmission model is the microscopic model of choice in disease dynamics and predator-prey models. Thus, we have two physically motivated endpoints — transmission-driven epidemic spread versus coherent environmental response — rather than the two endpoints of the interpolating family. Note that a variant of Eq. \ref{eq:newMicro} whose coupling is linear in the harmonic coordinate $u = 1/x$, rather than in the log coordinate, would close in the aggregate under the harmonic mean, by the same mechanism that closes the Gompertz model in the log domain, but that observable is unrealistic to record.

Further, since closure under aggregation is a property of the deterministic vector field, we measure the macroscopic closure error on the noise-free dynamics ($\sigma = 0$); this isolates the closure property from stochastic fluctuations and makes the comparison independent of the noise structure of Eq. \ref{eq:newMicro}. On real (non-mean-field) networks this closure error is consistently larger for the logistic model than for the Gompertz model (Figure \ref{fig:closure_error_comparison}), confirming the theoretical results. To quantify robustness, we run an ensemble of $20$ independent network and initial-condition realizations per topology, with both models sharing the same initial condition within each realization, and report the median relative error together with its interquartile ($25$--$75$th percentile) band. For each realization we summarize each model by its relative closure error averaged over the growth trajectory, $\bar{\epsilon}_m = \langle |\epsilon_m / \dot{X}_{\text{theory}}| \rangle_{0 < X < 1}$ for $m \in \{G, L\}$, and take the reduction factor to be the ratio $\bar{\epsilon}_L / \bar{\epsilon}_G$ of the logistic to the Gompertz value. Summarized as the median over the $20$ realizations, the Gompertz closure error is smaller than the logistic error by a factor of approximately $6$ on the heterogeneous Scale-Free network (interquartile range $5$--$14$) and approximately $12$ on the more homogeneous Erdős-Rényi network (interquartile range $9$--$15$); the Gompertz advantage holds in every realization of both topologies, with the heavier-tailed spread on the Scale-Free network reflecting the influence of high-degree hubs. The logistic closure error reflects the network-state correlations ($\text{Cov}_W(\mathbf{x}, \mathbf{x}) > 0$) inherent to transmission dynamics, while the Gompertz error arises from column-sum deviations that are self-correcting as nodes synchronize (both errors decrease as the system approaches saturation), as well as machine-precision inaccuracies in taking the natural logarithm of numbers close to 1. We omit the discrepancy analysis for the mean-field approximated networks where the Gompertz closure is exact under the doubly-stochastic scenario. Adding stochastic forcing consistent with Eq.\ \ref{eq:newMicro} (multiplicative in abundance, additive in the log-domain) leaves this closure property unchanged, but near saturation, where the theoretical growth rate approaches zero, the relative error becomes noise-dominated for both models; we therefore report the deterministic closure error, which cleanly isolates the aggregation property.

\begin{figure}[H]
  \centering
  \includegraphics[width=15.4cm]{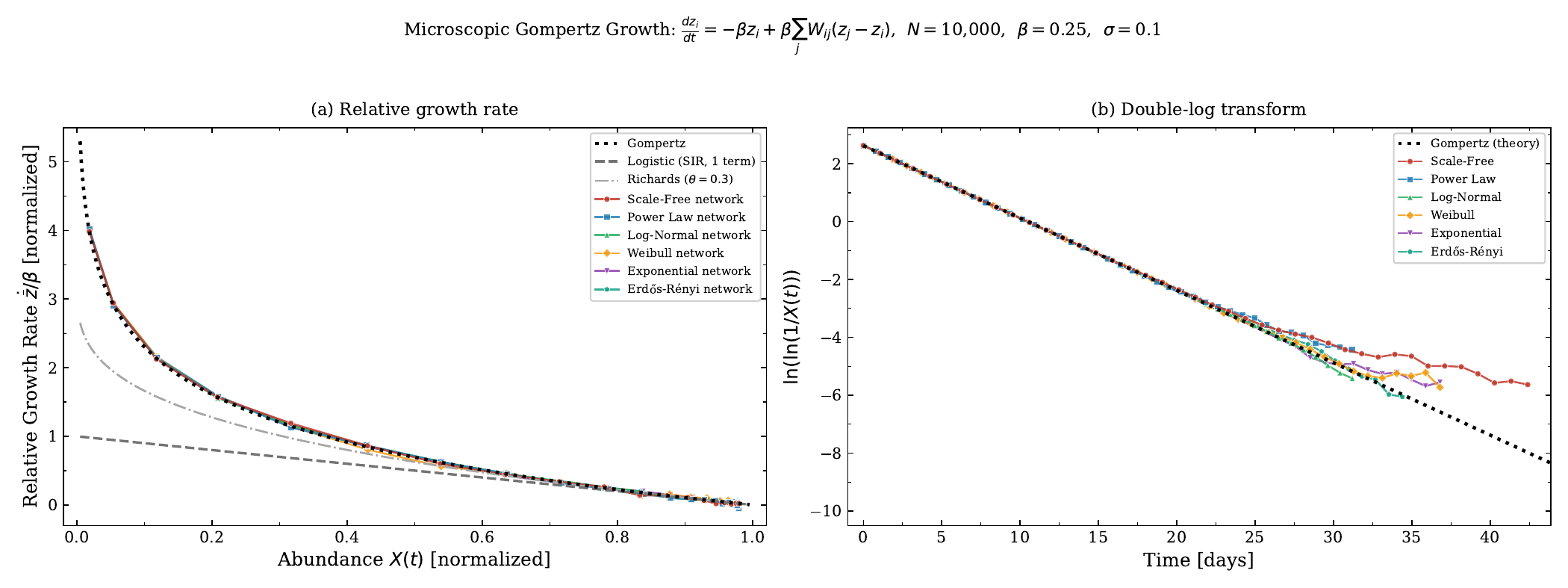}
  \caption{Comparison of relative growth rates for different networks under a microscopic Gompertz mode with 10,000 nodes using a mean field approximation, showing fidelity to the theoretical limit regardless of the underlying network. Different networks plotted in color, showing minimal deviation from the theoretical Gompertz curve. A logistic and Richards $\theta=0.3$ theoretical curve is plotted for comparison. (a) Relative growth plotted against abundance and (b) Double-log transform $\ln(\ln(1/X(t)))$ versus time, where a straight line indicates Gompertz growth. The abundance variable is normalized to values between 0 and 1.}
  \label{fig:gompertz_networks}
\end{figure} 

\section{Discussion and Conclusion}
\label{sec:discussion}

In conclusion, we have shown that Gompertzian growth is a maximally coherent growth process within a family of models with a power mean in the natural log-coordinate. We further expose specific mathematical properties for this model in this natural log-coordinate: 1. closure under aggregation, 2. stability in spite of internal fluctuation, 3. mean-variance uncoupling, and 4. linear neighbor contributions to the relative growth at the node-level, while local network effects vanish at the global level as if independent in line with the Gumbel correspondence. All of these properties shed new light on dynamics exhibited at the macroscopic level.

We can now further appreciate the Richards parametrization as a transition from non-collaborative logistic growth to maximally collaborative Gompertz growth as $\theta \rightarrow 0$.
This feature of $\theta$ was also obtained by Petroni et al. (2020) \cite{petroni2020logistic} by interpreting the $\theta-$logistic growth rate as non-linear resource availability dependent on the overall abundance, with linearity reached at $\theta=1$, without resorting to microscopic modelling. Coherence was there defined as maximally synergistic or collaborating elementary components. Used in a mortality model, the maximally coherent scenario implies that all units in the network are exposed to some external stressor simultaneously.
Molski and Konarski (2003) \cite{molski2003coherent} made a similar observation that Gompertzian growth is the coherent state in a quantum mechanical system with a time-dependent potential, an interpretation which sheds further light on the temporal nature of the postulated stress response.
This quantum mechanical system has also been used to describe coherent energy states of diatomic molecules in space \cite{morse1929diatomic}. 
In the proposed microscopic Gompertz model, the spectral gap of the deviation dynamics $\dot{\delta}=\beta(-2I + W)\delta$ protects the macroscopic state from microscopic perturbations through a structural property of the dynamics.

Two results of this paper bear on the source of growth in the Gompertz limit. First, the collective forcing term is purely redistributive: it exchanges growth between entities but contributes exactly zero to the growth of the aggregate (Eq. \ref{eq:redistributive}), which is governed entirely by each entity's own drift. Second, the identical macroscopic curve arises from a construction with no coupling whatsoever, since the normalized Gompertz curve is the Gumbel cumulative distribution and therefore the law obtained when each entity's conversion time is the extreme of many independent sub-processes.

Taken together, these say that Gompertz growth does not require a transmission network, and moreover that it cannot distinguish its existence from its non-existence from the macroscopic lens. A system of genuinely coupled entities and a collection of independent entities responding to a common driver produce the same aggregate trajectory. What the Gompertz form implies is therefore the existence of a common quantity to which entities respond.

Several existing models are instances of the common-driver reading. \citet{sacher1956statistical} and \citet{de2014stochastic} take the source of growth to be a stochastic perturbation; while De Lauro et al. do not discuss whether that perturbation is external, their formulation does not preclude it. \citet{markov2019reaction} recasts Gompertz growth in reaction-network terms, where growth is driven by a catalytic agent available to all reactants to the same degree — a ubiquitous driver in exactly the sense above. \citet{shklovskii2005simple} provides the extreme-value instance: mortality as the first escape past immune clearance, with the required exponential tail supplied by the Poisson zero-encounter probability, which gives way for the connection to the Gumbel distribution of extremes. That these mechanisms differ substantially in their physical content while converging on the same growth law is consistent with our finding that the Gompertz limit is precisely the regime in which such distinctions become invisible in the aggregate.

For many growth processes within biological organisms, it has been well established that the mediator can be the electromagnetic field \cite{becker1984electromagnetic}, which suggests that the accompanying growth pattern in these processes follows the Gompertz model as the result of a ubiquitous growth-inducing field. For growth in larger ecological or societal systems other catalysts can be air pollution, oxygen deprivation, etc. 

We finally note that Gompertz's original formulation concerned mortality rates as a function of age, where he observed geometric increase with arithmetic increase in age. It is indeed curious that we do not see mortality rates behave as linearly dependent on time which would be Gaussian, nor do we see any finite-ordered polynomial time dependence on mortality rates. Instead we see growth of mortality rates as induced by the accumulated state of the organism, acting through the endogenous state of the system itself.

\section{Methods}

\subsection{Running growth model simulations on networks}

The simulations of growth in a theoretical network structure are done using Equation \ref{eq:newMicro} with and without a mean-field approximation, using the shorthand notation $\dot{\bar{z}}=\frac{1}{N}\sum_i\frac{dz_i}{dt}$. In the mean-field case, each node's local neighbor-average is replaced by a degree-weighted global average. Specifically, the coupling target $\bar{z}_{\theta,i} = \frac{1}{\theta}\ln(\sum_j w_{ij} e^{\theta z_j})$ is approximated as $\bar{z}_\theta = \frac{1}{\theta}\ln(\sum_j w_j e^{\theta z_j})$ with weights $w_j = k_j / \sum_k k_k$, where $k_j = \sum_i a_{ij}$ is the degree of node $j$ and $a_{ij}$ is the adjacency matrix. This approximation is standard for configuration-model random graphs and becomes exact for large $N$ when the correlation between node state $z_j$ and degree $k_j$ is weak.

In the explicit network case, no such approximation is made. A configuration-model random graph is constructed from the degree sequence by creating a stub list (each node $i$ contributing $k_i$ stubs), randomly pairing stubs to form edges, and discarding self-loops; if the degree sum is odd, one randomly chosen node has its degree incremented by one so that the stubs pair exactly. Multi-edges are retained, so $a_{ij}$ carries the multiplicity of the edge between $i$ and $j$ and $k_i$ is the realized degree. The resulting adjacency structure is stored as a sparse neighbor list. The coupling weights are row-normalized as $w_{ij} = a_{ij}/k_i$, so that each node averages over its actual neighbors; a node left isolated after self-loop removal is assigned $w_{ii} = 1$, so its neighbor average equals its own state and the coupling term vanishes for that node. This normalization is not doubly stochastic for heterogeneous degree distributions, meaning the column sums $c_j = \sum_i w_{ij} = \sum_{i \in \mathcal{N}(j)} 1/k_i$ vary across nodes. As shown in the text, this does not materially affect the macroscopic Gompertz dynamics because the column-sum deviations are self-correcting under the stable coupling.

We simulate $N = 10{,}000$ nodes with parameters $\beta = 0.25$, $\sigma = 0.1$, and time step $\Delta t = 0.02$; these values, together with the noise scaling and the initial-condition spread, are held identical across the stochastic growth-sweep experiments so that differences between figures reflect the dynamics rather than differing numerical choices (the closure-error diagnostic below is instead evaluated on the deterministic dynamics, $\sigma = 0$, as explained there). Initial conditions for the growth-sweep experiments are drawn as $z_i \sim \mathcal{N}(-13.8, \, 0.3^2)$, corresponding to initial abundance $x_i \approx 10^{-6}$. Integration is performed using the Euler-Maruyama scheme with noise increments $\sigma\sqrt{\Delta t}\, \xi_i$, where $\xi_i \sim \mathcal{N}(0,1)$, consistent with the $\sigma\,dB_i$ convention of Eq.\ \ref{eq:newMicro}. All simulations use a fixed base random seed for reproducibility, and the quantitative closure comparison below is additionally averaged over an ensemble of seeds. For $\theta$ near $1$, the mean-variance coupling identified in Eq.\ \ref{eq:meanVarianceCoupling} produces visible noise in the later stages of the simulations.

The node degrees are drawn from the following distributions, each representing a distinct network topology:

\begin{itemize}
\item \textbf{Scale-free}: $P(k) \propto k^{-3}$, $k_{\min} = 2$, generated via inverse CDF sampling, approximating a Barab\'{a}si-Albert network.
\item \textbf{Power law}: $P(k) \propto k^{-2.2}$, $k_{\min} = 2$, a heavier-tailed variant generated via inverse CDF sampling.
\item \textbf{Log-normal}: $\ln k \sim \mathcal{N}(1.5, \, 0.8^2)$, $k_{\min} = 2$.
\item \textbf{Weibull}: $k \sim \text{Weibull}(\text{shape}=1.5, \, \text{scale}=5)$, $k_{\min} = 2$.
\item \textbf{Exponential}: $k \sim \text{Exp}(\text{scale}=4)$, $k_{\min} = 2$.
\item \textbf{Erd\H{o}s-R\'{e}nyi}: $k \sim \text{Poisson}(\lambda = 6)$, $k_{\min} = 2$.
\end{itemize}

For the Richards interpolation experiments, the coherence parameter $\theta$ is varied in $\{0, 0.1, 0.3, 0.5, 0.7, 1.0\}$ on a single network (log-normal degree distribution). Since logistic-like dynamics ($\theta \to 1$) exhibit slower growth from low initial abundance, the simulation duration is extended as $T_{\max} = 50 \cdot \max(1, 1 + 4\theta)$ to ensure all trajectories reach saturation. The macroscopic observable is the geometric mean $X(t) = \exp(\frac{1}{N}\sum_i z_i)$, and the relative growth rate is computed numerically as $\gamma(t) = \dot{X}(t) / X(t)$ using a central difference scheme for $\dot{X}$.

\subsection{Measuring macroscopic closure error}

To quantify the degree to which each model's macroscopic equation holds on real (non-mean-field) networks, we compute the relative closure error for both Gompertz and logistic dynamics. For the Gompertz model, the closure error is defined as
\begin{equation}
\epsilon_G(t) = \dot{\bar{z}}(t) - (-\beta\bar{z}(t)),
\end{equation}
where $\dot{\bar{z}} = \frac{1}{N}\sum_i \dot{z}_i$ is the actual macroscopic rate of change computed from the microscopic dynamics, and $-\beta\bar{z}$ is the theoretical closed-form prediction. For the logistic model, the microscopic dynamics are $\dot{x}_i = \beta(1-x_i)\sum_j w_{ij}x_j$, and the closure error is
\begin{equation}
\epsilon_L(t) = \dot{\bar{x}}(t) - \beta\bar{x}(t)(1-\bar{x}(t)),
\end{equation}
where $\dot{\bar{x}} = \frac{1}{N}\sum_i \dot{x}_i$ and $\beta\bar{x}(1-\bar{x})$ is the macroscopic logistic prediction. The relative closure error is then $|\epsilon / \dot{X}_{\text{theory}}|$, which gives a dimensionless measure of how much the actual macroscopic dynamics deviate from the closed-form equation, normalized by the expected growth rate.

Both models are simulated on explicit configuration-model networks with row-normalized coupling $w_{ij} = a_{ij}/k_i$, using the same initial conditions ($z_i \sim \mathcal{N}(-2.3, 0.3^2)$, corresponding to $x_i \approx 0.1$) so that both models grow through the full trajectory during the simulation window. This is the only experiment that uses a higher initial abundance than the growth-sweep experiments, because the logistic model, whose growth is proportional to abundance, requires a non-negligible seed to grow at all; within the experiment the Gompertz and logistic models share the exact same initial condition. The relative closure error is averaged over the whole growth trajectory ($0 < X < 1$); because the measurement is deterministic it stays finite throughout (both models' errors decay to zero at saturation), so no restriction to a particular growth window is required. Because closure under aggregation is a property of the deterministic vector field, we measure the closure error on the noise-free dynamics ($\sigma = 0$). This isolates the aggregation property and removes any dependence on the noise structure, which is important because the microscopic noise of Eq.\ \ref{eq:newMicro} is additive in the log-domain (equivalently, multiplicative in abundance) and is therefore not symmetric between the log-domain Gompertz model and the abundance-domain logistic model. To quantify reproducibility rather than reporting single draws, we repeat the experiment over an ensemble of $20$ independent network and initial-condition realizations and report the median relative error with its interquartile ($25$--$75$th percentile) band; a smoothing window of $25$ time steps is applied to each realization's relative error for visual clarity. We verified that adding stochastic forcing consistent with Eq.\ \ref{eq:newMicro} leaves the deterministic closure property unchanged, but that near saturation the theoretical growth rate vanishes and the \emph{relative} error there becomes dominated by the noise for both models (with the direction of the residual difference depending on the assumed noise structure); the deterministic closure error avoids this confound and is the quantity reported. Parameters: $N = 10{,}000$, $\beta = 0.25$, $\Delta t = 0.02$, $T_{\max} = 50$, $\sigma = 0$ (deterministic).

\section*{Competing interests}
The authors declare no competing interests.

\appendix
\section*{Code availability}
Code is available at [https://github.com/matzhaugen/GompertzManuscript]
\addcontentsline{toc}{section}{Appendices}
\renewcommand{\thesubsection}{\Alph{subsection}}

\section*{Appendix A: Time-sensitivity analysis}

\begin{figure}[H]
  \centering
  \includegraphics[width=15.4cm]{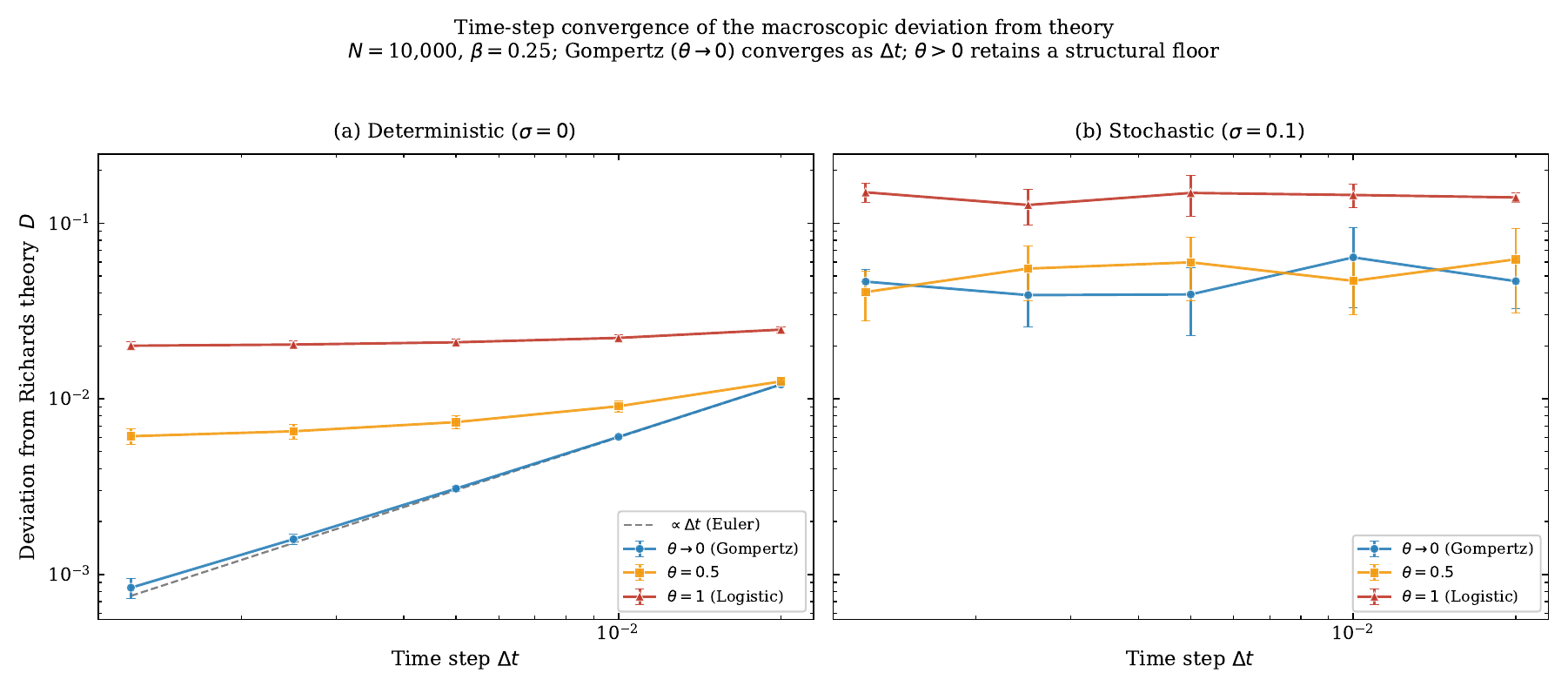}
  \caption{Time-step ($\Delta t$) convergence of the deviation between the simulated macroscopic trajectory and the closed-form Richards solution, quantifying whether the high-$\theta$ deviations of Figure \ref{fig:gompertz_richards_interpolation} survive the continuum limit. $D$ is the root-mean-square difference, in the double-log domain $\ln(\ln(1/X))$, between simulation and theory over the growth trajectory ($0.1 \le X \le 0.99$), averaged over independent initial-condition realizations ($N = 10{,}000$ nodes, log-normal network, $\beta = 0.25$). (a) Deterministic dynamics ($\sigma = 0$): the Gompertz control ($\theta \to 0$) follows the first-order Euler reference $\propto \Delta t$ (grey dashed) and vanishes as $\Delta t \to 0$, whereas $\theta > 0$ converges to a nonzero structural floor that grows with $\theta$ (the mean-variance coupling of Eq. \ref{eq:meanVarianceCoupling}). (b) Stochastic dynamics ($\sigma = 0.1$): the deviations are essentially independent of $\Delta t$ and grow with $\theta$, showing that the near-saturation fluctuations are genuine stochastic features rather than discretization error. Error bars are $\pm 1$ standard deviation across realizations.}
  \label{fig:convergence_deltat}
\end{figure}

To confirm that these high-$\theta$ deviations are a genuine structural effect and not an artifact of Euler--Maruyama discretization, we performed a time-step convergence study (Figure \ref{fig:convergence_deltat}), measuring the root-mean-square deviation $D$ of the simulated macroscopic trajectory from the closed-form Richards solution as $\Delta t$ is refined. We repeat the $\theta \in \{0, 0.5, 1\}$ simulations at time steps $\Delta t \in \{0.02, 0.01, 0.005, 0.0025, 0.00125\}$, averaging $D$ over $3$ independent initial-condition realizations in the deterministic case and $6$ in the stochastic case. The integration horizons are $T_{\max} = 50$, $80$ and $110$ for $\theta = 0$, $0.5$ and $1$ respectively, each chosen so that the trajectory reaches saturation within the comparison window; these are shorter than the $T_{\max} = 50 \cdot \max(1, 1 + 4\theta)$ used for the growth sweep of Figure \ref{fig:gompertz_richards_interpolation}, which is sufficient here because the deviation is measured only over $0.1 \le X \le 0.99$. The reference trajectory is the exact closed-form macroscopic solution: substituting $y = X^{\theta}$ maps the $\theta$-logistic macroscopic equation to the logistic equation $\dot{y} = \beta y(1-y)$, giving $X_{\text{theory}}(t) = \left[y_0/\left(y_0 + (1-y_0)e^{-\beta t}\right)\right]^{1/\theta}$ with $y_0 = X(0)^{\theta}$, and $X_{\text{theory}}(t) = \exp\!\left(\ln X_0 \, e^{-\beta t}\right)$ in the Gompertz limit $\theta \to 0$. The Gompertz limit ($\theta \to 0$) serves as a control: because it is exactly closed under aggregation, its deviation must vanish in the continuum limit, and indeed $D$ decreases in proportion to $\Delta t$ (first-order Euler convergence) toward zero. For $\theta > 0$, by contrast, $D$ converges to a nonzero floor that grows with $\theta$ (from $\approx 0.6\%$ at $\theta = 0.5$ to $\approx 2\%$ at $\theta = 1$ in the deterministic case), demonstrating that a genuine structural deviation — the mean-variance coupling of Eq. \ref{eq:meanVarianceCoupling} — survives $\Delta t \to 0$ and is cleanly separated from the (vanishing) discretization error. When noise is included, the deviations become essentially independent of $\Delta t$ and grow with $\theta$, confirming that the near-saturation fluctuations amplified by the double-log transform are genuine stochastic features of the dynamics rather than integration error. The residual numerical component present at coarse $\Delta t$ can, if desired, be reduced by refining the step or by adaptive sub-stepping, but the structural $\theta$-dependent effect cannot.

\bibliographystyle{abbrvnat}
\bibliography{references2}  






\end{document}